\documentclass[oldversion, preprint]{aa}

\usepackage{graphicx}
\usepackage{txfonts} \usepackage{longtable} \usepackage{multirow}
\begin{document}

\title{AMBER observations of the AGB star RS Cap: Extended atmosphere and
comparison with stellar models\thanks{Based on observations made with ESO 
Telescopes at the Paranal Observatory under programme ID 383.D-0619}}

\titlerunning{AMBER observations of RS Cap}

   \author{I. Mart\'i-Vidal\inst{1,2} 
\and J.M. Marcaide\inst{1} \and A. Quirrenbach\inst{3} \and K. Ohnaka\inst{2} 
\and J.C. Guirado\inst{1} \and M. Wittkowski\inst{4} 
}

   \institute{Dpt. Astronomia i Astrof\'isica, Universitat de Val\`encia, C/ Dr.
Moliner 50, E-46100 Burjassot (Spain)\\ \email{imartiv@mpifr-bonn.mpg.de} \and
Max-Planck-Institut f\"ur Radioastronomie, Auf dem H\"ugel 69, D-53121 Bonn
(Germany) \and Universit\"at Heidelberg, Landessternwarte K\"onigstuhl 12,
D-69117 Heidelberg (Germany) \and European Southern Observatory,
Karl-Schwarzschild-Str. 2, D-85748 Garching bei M\"unchen (Germany)} 

\date{Accepted for publication in A\&A}

\abstract{
We report on K-band VLTI/AMBER observations at medium spectral resolution
($\sim$1500) of RS Capricorni, an M6/M7III semi-regular AGB star. From the
spectrally dispersed visibilities, we measure the star diameter as a function of
observing wavelength from 2.13 to 2.47 microns. We derive a Rosseland angular 
diameter of $7.95 \pm 0.07$ mas, which corresponds
to an effective temperature of $3160 \pm 160$ K. We detect size variations of
around 10\% in the CO band heads, which are indicative of strong opacity effects of CO
in the stellar photosphere.  We also detect a
linear increase in the size as a function of wavelength, beginning at
2.29 microns. Models of the stellar atmosphere, based on the mass of the
star estimated from stellar-evolution models, predict CO-size effects of about
half the sizes observed, and cannot reproduce the linear size increase 
with wavelength, redward of 2.29 microns. We are able to model 
this linear size increase with the addition of an extended water-vapor envelope
around the star. However, we are unable to fit the data in the CO bandheads. 
Either the mass of the star is overestimated by the stellar-evolution models 
and/or there is an additional extended CO envelope in the outer part of the 
atmosphere. In any case, neither the water-vapor envelope nor the CO 
envelope can be explained using the current models.}

 \maketitle

\section{Introduction}

The effective size of a star at a given wavelength depends on the opacity of the
stellar atmosphere at that wavelength.  Since we effectively measure the
diameter of the $\tau = 1$ surface of the star, the size is related to the
extension of the atmospheric region where the absorption is produced.  The
diameter of that surface varies with opacity, hence frequency. The
atmospheres of cool giants are so extended that these size variations are
observable with an interferometer. Quirrenbach et al.  (\cite{Quirrenbach1993}, 
\cite{Quirrenbach2001})
studied the extended absorption regions of stellar atmospheres of cool giant
stars in the TiO band at 712\,nm with the MkIII interferometer. These authors
measured the visibilities of a set of 47 stars using two filters, one centered
in the TiO band, and the other in the continuum part of the spectrum close to
the TiO band (754\,nm).  They found that the sizes in the TiO band are
larger than those in the continuum, and that this effect is stronger for cooler
stars. The sizes in the TiO absorption band are $\sim 10$\% larger than in the
continuum for $R-I$ color indices of $\sim1.6$ (spectral types M3--M4) and as 
much as 30\% larger for $R-I$ color indices of $\sim2.2$ (spectral types M6--M7).

Similar differences between the sizes observed in the continuum and in bands 
containing absorption band heads of other molecules (such as H$_2$O or CO) have 
been reported for AGB stars. Mennesson et al. (\cite{Mennesson2002}) and Perrin 
et al. (\cite{Perrin2004}), for instance, reported very large size ratios 
(50\% or more) for some stars.

Quirrenbach et al. (\cite{Quirrenbach1993}, \cite{Quirrenbach2001}) were successful 
in qualitatively
reproducing their data with the latest set of cool giant models from the
general-purpose stellar atmosphere code {\tt PHOENIX}. Spherical, hydrostatic,
massively line-blanketed atmosphere models were constructed and used to predict
the uniform-disk diameters in the TiO band and the continuum band as a
function of model effective temperature, surface gravity, and mass (the stellar
mass was used in the modeling, since it controls the deviation from
plane-parallel atmospheres.) For most of the observed oxygen-rich giants, the
diameter ratios of the TiO band to the continuum band agreed with the
models computed for masses $\sim 0.5\,M_\odot$. Hence, model atmospheres with very 
low stellar masses could fit the large diameter ratios observed in many stars, 
although evolutionary models predict masses as high as 5\,$M_\odot$. 

A possible explanation of this inconsistency would be the existence of a
transition zone at the base of the stellar wind (the {\em MOLsphere}; e.g. 
Tsuji \cite{Tsuji2008}, and references therein), which could provide sufficient
opacity in the TiO bands (and other molecular bands) to make AGB stars appear 
much larger than predicted by the hydrostatic model atmospheres such as {\tt PHOENIX}. 
According to this picture, one would expect similar
size effects for the CO band heads. Therefore, we decided to use AMBER in the K
band to measure the effective sizes of a set of four cool giant stars through
the CO band heads at 2.3\,$\mu$m.  The use of AMBER in medium-resolution mode
($\lambda/\Delta\lambda \sim 1500$) provides considerably more information than 
could be obtained with the MkIII
interferometer (Quirrenbach et al. \cite{Quirrenbach1993}) and the IOTA 
interferometer (Mennesson et al. \cite{Mennesson2002}; Perrin et al. \cite{Perrin2004}), 
in which narrow-band filters were used. In this paper, we 
report on the
results obtained from the analysis of the observations of the first star of our
sample: RS Cap. 

RS Cap (HD\,200994) has a K-band magnitude of $-0.2$ (Cutri et al. \cite{cutri2003}).
It is a semi-regular variable (SRb) of spectral type M6/M7III 
and is located at $\alpha = 21\textrm{h}\,07\textrm{m}\,15.4\textrm{s}$, $\delta =
-16^{\circ}\,25'\,21.4''$ (J2000.0). It has a visual magnitude of 8.3 and a
parallax $\pi = 1.26\pm0.86$\,mas (van Leeuwen \cite{Leeuwen2007}), which 
maps into a distance between $500$ and $2500$\,pc.
However, following Scalo (\cite{Scalo1976}) or Winters et al. (\cite{winters03}), 
a bolometric absolute magnitude between $-4$ and $-5$ is derived for RS Cap (around 
$-4.8$ in the case of Winters et al.). These estimates, in addition to the spectrum 
fitted between 0.1 and 2$\mu$m, translate into a distance of $\sim$300\,pc (Richichi et 
al. \cite{Richichi1992}). 

A variability amplitude
of $\Delta V \sim 0.5$ is seen in the Hipparcos data (Perryman
\& ESA \cite{Perryman1997}), although a larger variability amplitude (in photometric 
magnitude) of $\Delta B \sim 2$ was reported in Kukarkin et al. (\cite{Kukarkin1969}),
which has a period of $\sim340$\,days.

The remainder of this paper is structured as follows: In Sect.
\ref{OBSERVATIONS}, we describe our AMBER observations and the strategy followed
in the data calibration and reduction. In Sect. \ref{RESULTS}, we report on the
results obtained: Sect. \ref{TEFF} is centered on our estimate of the effective
temperature and Sect. \ref{CObands} on the size effects observed at the CO band heads.
In Sect. \ref{MODELING}, we compare our results with synthetic data obtained from model 
stellar atmospheres. In Sect. \ref{CONCLUSIONS}, we summarize our conclusions.

\section{Observations and data reduction} \label{OBSERVATIONS}

We observed RS Cap with the ESO Very Large Telescope Interferometer (VLTI)
using the Astronomical Multi-BEam combineR, AMBER (see Petrov et al.
\cite{Petrov2007} for details on this instrument), in medium-resolution mode
($\lambda/\Delta\lambda \sim 1500$). This instrument performs
simultaneous observations of the interferometric fringes generated by three
telescopes. Therefore, it measures {\em closure phases}, which are
quantities independent of atmospheric or instrumental telescope-dependent
contributions (e.g. Rogers et al. \cite{Rogers1974}). AMBER also measures the 
so-called {\em differential phase} on each baseline, which roughly represents the 
phase of spectral features with respect to that in the continuum\footnote{To be 
precise, two pieces of information are lost in the differential phase, compared to the 
original Fourier phase: the phase offset (or phase at the first spectral channel) and 
the phase gradient (phase as a linear function of wavenumber).}. The observations were
performed on 4 June 2009, from 06:30\,UT to 08:30\,UT, using the three VLTI
Auxiliary Telescopes (AT) positioned on stations E0, G0, and H0. These stations
are distributed roughly in the east-west direction. In Table \ref{OBSERVCONFIG},
we give the projected baseline lengths and position angles for our observations,
together with a summary of the atmospheric observing conditions. 18 Cap
(M0III star located at $\alpha = 20\textrm{h}\,51\textrm{m}\,49.3\textrm{s}$,
$\delta = -26^{\circ}\,55'\,08.9''$, J2000.0) was also observed as a calibrator,
with the same observing configuration used to obtain the RS Cap visibilities.

\begin{table*} 

\centering 

\begin{tabular}{ c | c | c c c | c c c | c } 

\hline\hline

& {\bf UTC} & {\bf Seeing} & {\bf Air mass} & {\bf Coh. time} & {\bf E0-G0} & {\bf G0-H0} &
{\bf E0-H0} & {\bf PA}\\ 

& (hh:mm) & ($''$) &  & (ms) & L(m) & L(m) & L (m) & (deg)\\

\hline RS Cap & 07:08--08:06 &  0.61 / 0.94  &  1.03 / 1.12  &  1.6 / 2.4  & 14.4 / 15.5 & 28.7
/ 31.0 & 43.1 / 46.5 & 119 / 113 \\ 

18 Cap & 06:47--06:56 / 08:16 -- 08:26 & 0.57 / 1.19  &  1.00 / 1.11  &
1.5 / 2.7  & 14.9 / 16.0 & 29.7 / 31.9 & 44.6 / 47.9 & 110 / 123 \\ 

\hline

\end{tabular} 

\caption{Details of our observations. For each star, we give the
ranges of values obtained during all the exposures. L(m) is the projected
baseline length (in meters) and PA is the position angle of the baseline (North
through East). Coh. time is the coherence time estimated from the atmospheric
conditions.} 

\label{OBSERVCONFIG} 

\end{table*}

The high flux densities of RS Cap and 18 Cap in the H band (magnitudes of 0.25
and 0.42, respectively), allowed us to use the fringe tracker FINITO (see Gai et
al. \cite{Gai2004}), which employs part of the H-band light to correct, in real
time, the delay shifts of the fringes due to atmospheric turbulences, thus
increasing the coherence of the signal. A larger coherence allows the use of a
longer detector integration time (DIT).  Given the moderately good atmospheric
conditions during our observations (see Table \ref{OBSERVCONFIG}), we achieved a
DIT of 200\,ms with a negligible signal loss due to atmospheric jitter. This allowed
us to simultaneously observe a large spectral range of the K-band (between 2.13
and 2.47\,$\mu$m) using medium-resolution mode. This wavelength range contains
the $^{12}$CO (2$-$0), (3$-$1), and (4$-$2), as well as the $^{13}$CO (2$-$0)
band heads, together with the continuum blueward of the $^{12}$CO (2$-$0) band
head at 2.29\,$\mu$m.

The data acquisition was carried out in the following order: (1) observation of an
artificial signal for calibration of the instrumental dispersive effects; (2) one
exposure of dark frames (an exposure consisted on 200 frames of 200\,ms each); (3) 
five exposures of the target; and (4) one exposure of empty sky close to the target. The
observations of dark, target, and sky exposures were iterated once 
for 18\,Cap (spanning 10 minutes), three times for RS\,Cap (spanning 1 hour),
and one more time for 18\,Cap (spanning 10 minutes). 
The dead time between iterations, when the sources were not observed, was dedicated 
to the setup of the instrument: preparation of the optical-delay lines, telescope
pointing, etc.

The raw visibilities and differential closure phases for RS Cap and 18 Cap were 
obtained using the {\tt amdlib} libraries (version 2.2)
and the interface provided by the Jean-Marie Mariotti Center (JMMC). First, we
manually aligned the spectrally-dispersed photometry channels of the ATs to the
interferometric channel, based on observations of the target and calibrator
stars. We removed the bad pixels and took into account the flat, dark, and sky
contributions. Afterwards, we calibrated the
instrumental dispersive effects and fringe-fitted each frame of RS Cap and 18
Cap (based on the P2VM algorithm of Tatulli et al. \cite{Tatulli2007}). 
The resulting visibilities of individual frames were selected and
averaged for each exposure in several ways, to test the robustness of our
results on different averaging and selection schemes.  The finally chosen
selection scheme was based on an atmospheric piston criterion (keeping only
frames with a piston smaller than 8\,$\mu$m, to select only well-detected and
centered fringes in all three baselines) and keep 50\% of the remaining
frames based on a signal-to-noise ratio (SNR) criterion.

After the frame selection, we obtained a single frame-averaged visibility spectrum 
for each exposure. The last step in the data reduction was to
average the visibilities of all exposures of each star to increase the SNR. This
step was performed outside {\tt amdlib}, with an in-house developed python-based
program that uses the PyFITS library (provided by the Space Telescope Science
Institute, operated by AURA for NASA). The averaged closure phases and 
differential phases (i.e., those
resulting from the averaged exposures) are zero within $3^{\circ}$, in
absolute value, for both stars and for all the spectral channels. These small
values of the closure phases indicate that the emission from both sources is
symmetric, at least in the direction of the projected baselines of our
observations and to the scale corresponding to our spatial resolution.
This is unsuprising, given that we probe only the first lobe of the Fourier
transform of the observed source structure, which is, therefore, only 
partially resolved.

The uncertainties in the amplitude and phase for each spectral channel were computed
from their standard deviations during the averaging of the exposures. Fractional
uncertainties of only $1-2$\% in the observed visibilities were obtained for 
all channels and baselines.

\subsection{Visibility amplitude and wavelength calibration}
\label{LambdaCal}

A standard calibration procedure, to assign the corresponding
wavelength to each spectral channel of the AMBER detector, is not currently 
available. This is a concern,
especially for the mid-resolution and low-resolution modes, where it is
difficult, or impossible, to use line profiles either from the observed objects
or from Earth's atmosphere for the calibration. We performed the wavelength 
($\lambda$)
calibration of our AMBER observations by comparing the positions of the minima
of the observed CO band heads in RS Cap to their values derived from the models
described in Sect. \ref{RESRSCAP}. The use of different stellar models did not
affect our $\lambda$ calibration at a level higher than our spectral resolution.

In the amplitude calibration, the CHARM2 catalog of high
angular-resolution stellar-diameter measurements (Richichi, Percheron, \&
Khristoforova \cite{Richichi2005}) indicates that the uniform-disk K-band diameter of our
calibrator star, 18 Cap, is 5.02\,mas. This uniform-disk diameter was used to
calibrate the visibility amplitudes by comparing those of 18 Cap to the model
predictions corresponding to a uniform disk of 5.02\,mas diameter. This
amplitude calibration was performed as described in Annex B of the AMBER Data
Reduction Software User Manual (Revision 2.1) of the JMMC\footnote{{\tt
http://www.mariotti.fr/data\_processing\_amber.htm}}.

\subsection{Photometry calibration}

\begin{figure} 

\centering 

\includegraphics[width=9cm]{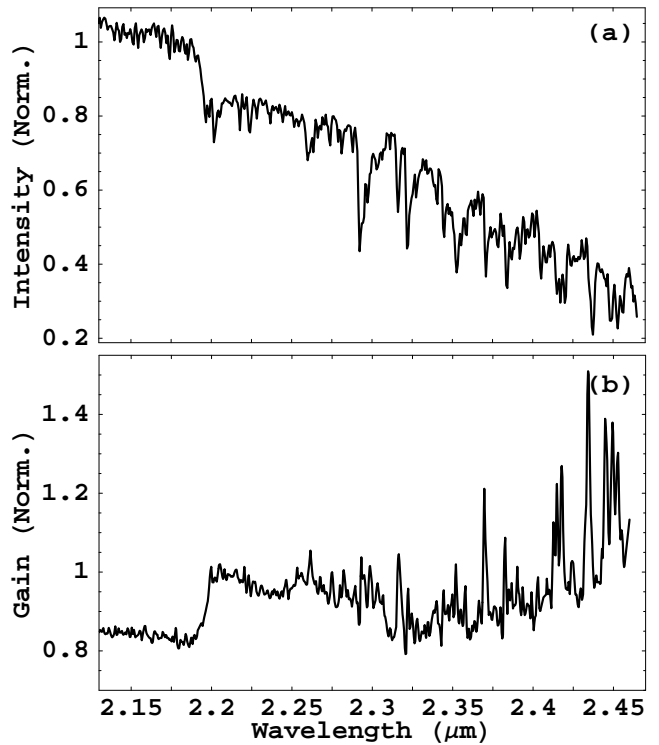}

\caption{(a) Normalized spectrum of RS Cap, as observed with the ATs; (b) inverse of 
the atmosphere$+$detector spectral transmission profile.}

\label{SpectrumFigure} 

\end{figure}

In Fig. \ref{SpectrumFigure}(a), we show the normalized spectrum of RS Cap, obtained
from the average of the photometry channels of the three ATs.
We note that there is a clear discontinuity in the spectrum at 
$\sim 2.2$\,$\mu$m (a similar discontinuity is also seen in the spectrum of 
18 Cap). This discontinuity is caused by the transmission of 
the K-band fiber of the AMBER instrument. Therefore,
it can be understood as an instrumental multiplicative effect, 
which biases the photometric and interferometric channels in the same way for wavelengths 
shorter than $\sim 2.2$\,$\mu$m. An additive contribution to these channels would have 
affected the contrast of the fringes relative to the photometry and, therefore, 
would have translated into a changing amplitude of the visibilities, which is not 
observed. 
Since any multiplicative bias in the data has no effect on the 
visibility amplitudes, the atmospheric opacity and any unmodeled gain effect in 
the detector do not affect the visibilities, but only the observed spectra. 

If we wish to use the observed spectrum of RS\,Cap in our 
modeling, it is necessary to calibrate it with the atmospheric (and instrumental) spectral 
transmission profiles. To perform this calibration, we proceeded in the following 
way: 1) we obtained a template spectrum of an M0III star, which is corrected for atmospheric 
effects (star BS4371, observed\footnote{This spectrum has a resolution of 
$\lambda/\Delta\lambda \sim 1100$, similar to that of our observations} by Lan\c{c}on et 
al. \cite{Lancon2000}); 
2) we divided the template spectrum by the observed spectrum of 18 Cap, 
to obtain the inverse of the transmission profile of the atmosphere plus the detector 
(we call this quantity {\em gain} and we show it in Fig. \ref{SpectrumFigure}(b)); and  
3) we calibrated the spectrum of RS Cap, multiplying it by the gain. The 
normalized calibrated spectrum is shown in Fig. \ref{RSCAPDISC}(a). By applying this 
calibration strategy, we assume that the airmass is similar for 
target and calibrator, which is indeed the case in our observations (see 
Table \ref{OBSERVCONFIG}).

We note that there are some atmospheric telluric lines that can be clearly 
seen in the transmission profile shown in Fig. \ref{SpectrumFigure}(b); those 
at 2.45 (three lines), 2.44, 2.42 (three lines blended), 2.37, and 2.32\,$\mu$m. 
The effect of the optical-fiber transmission jump at 2.2\,$\mu$m is also seen 
in the gain, together with a slight increase as a function of wavelength redward 
of $\sim$2.35\,$\mu$m due to water vapor in the Earth's atmosphere. We also note 
that, although the 
CO bands might have slightly different depths in 18\,Cap and BS4371, no clear 
features are seen in the regions of the CO bands for the gain shown in 
Fig. \ref{SpectrumFigure}(b).

\section{Results and discussion} 
\label{RESULTS}

\subsection{Continuum angular diameter and effective temperature} 
\label{TEFF}

\begin{figure} 

\centering 

\includegraphics[width=9cm]{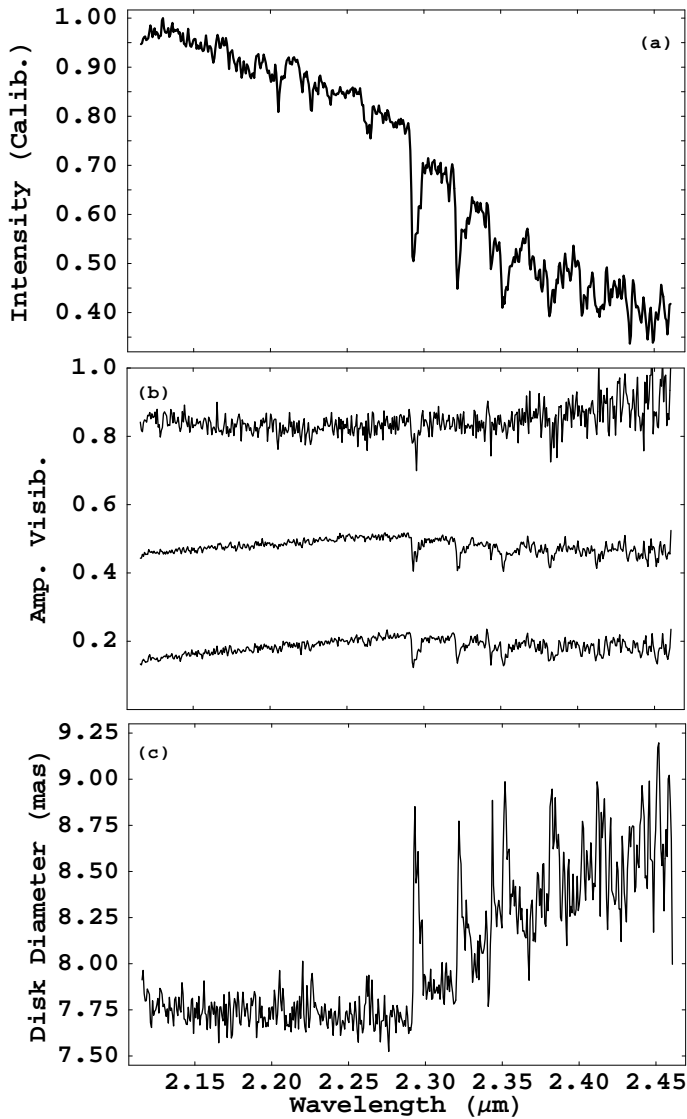}

\caption{(a) Spectrum of RS Cap (normalized); (b) calibrated visibility
amplitudes; (c) uniform-disc angular diameter at each AMBER spectral channel.}

\label{RSCAPDISC} 

\end{figure}

In Fig. \ref{RSCAPDISC}, we show (a) the normalized spectrum of RS Cap, (b) the 
amplitude visibilities for the three baselines (higher visibility corresponds to
shorter baseline), and (c) the resulting diameter estimates obtained from the
fit of a uniformly-bright disk to the visibilities at each spectral channel. 

The mean uniform-disk angular diameter fitted to the RS Cap visibilities in the
continuum (i.e., blueward of $2.29$\,$\mu$m) is $\theta = 7.75\pm0.07$\,mas.
To derive the effective temperature accurately, the size estimated with a 
limb-darkening model must be used, instead of that estimated from a uniform-disc 
model. We note that $T_{\mathrm{eff}}$ is a function of the criterion adopted for 
the radius of the star. In spherical atmospheres, the actual $T_{\mathrm{eff}}$ 
used to compare to the models should be computed using the diameter at which the 
Rosseland opacity is 1 (i.e., the Rosseland diameter, 
$\theta_{\mathrm{Ross}}$; see, e.g., Sect. 3.4 of Wittkowski, Aufdenberg, \& 
Kervella \cite{Witt2004}).

\begin{figure} 

\centering 

\includegraphics[width=9cm]{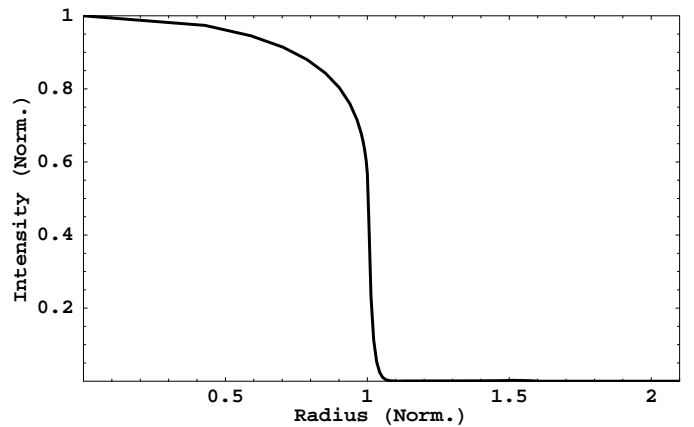}

\caption{Monochromatic intensity profile, normalized to $R_{\mathrm{Ross}} = 1$, 
corresponding to the {\tt MARCS} model described in Sect. \ref{RESRSCAP}. This profile
was computed in the continuum part of the spectral coverage of our observations.}

\label{LimbFig} 

\end{figure}

We estimated the Rosseland diameter of RS Cap by fitting the 
visibilities in the continuum part of our spectral coverage to the Hankel transform 
(i.e., Fourier transform of a circularly-symmetric source) of 
the monochromatic intensity profile of the {\tt MARCS} atmospheric model 
described in Sect. \ref{RESRSCAP}. We show this intensity profile, normalized to a 
Rosseland radius $R_{\mathrm{Ross}} = \theta_{\mathrm{Ross}}/2 = 1$, in Fig. 
\ref{LimbFig}. If this profile is 
given as $I(r)$ (where $I$ is the intensity normalized to 1 and $r$ is the 
radial coordinate normalized to the Rosseland radius, $R_{\mathrm{Ross}} = 1$), then the 
model used to fit the visibilities is 

\begin{equation}
V(q) = \int_0^\infty{I(r/R_{\mathrm{Ross}})J_0(K\,q\,r)\,r\,\mathrm{d}r},
\label{HankEq}
\end{equation}

\noindent where $J_0$ is the first-kind Bessel function of order 0, $q$ is the baseline 
length, and $K$ is a constant to scale $r$ and $q$ to their corresponding units (e.g., 
milliarcseconds and megawavelengths, respectively). The use of this fitting model, instead 
of the Hankel transform of a uniform disc, allows us to take into account limb-darkening and, 
simultaneously, to estimate directly the Rosseland radius 
($R_{\mathrm{Ross}}$ in Eq. \ref{HankEq}).

The best-fit Rosseland diameter of RS Cap is $\theta_{\mathrm{Ross}} = 7.95 \pm 0.07$\,mas. 
Our size estimate, together with the bolometric flux of 
$(2.1\pm0.2)\times10^{-6}$\,erg cm$^{-2}$ s$^{-1}$ reported in Richichi et al. 
(\cite{Richichi1992}), translates into an effective temperature of 
$T_{\mathrm{eff}} = 3160 \pm 160$\,K. 

Lunar occultations of RS Cap, recorded
with the TIRGO telescope at 2.2\,$\mu$m, resulted in a uniform-disk diameter estimate
of $7.75\pm0.67$\,mas (Richichi et al. \cite{Richichi1992}). This diameter, together with
the bolometric measurements performed by the same authors, implies an
effective temperature of $\sim 3200$\,K, higher than the temperature derived
from the calibration of Bessel, Castelli \& Plez (\cite{Bessel1998}) ($\sim
2900$\,K) and that derived from the calibration of Scalo (\cite{Scalo1976})
($\sim 2600$\,K), but in agreement with the calibration given in
Ridgway et al. (\cite{Ridgway1980}). A later re-analysis of the Lunar-occultation
events (Richichi et al.  \cite{Richichi1999}), to estimate the Rosseland angular diameter,
resulted in an even higher effective temperature for this star
($3481\pm177$\,K). Dyck, van Belle \& Thomson (\cite{Dyck1998})
estimated a size of $7.0 \pm 0.8$\,mas from a single-baseline interferometric
observation in the K band using the Infrared Optical Telescope Array (IOTA).
This size also translates into a higher effective temperature for RS Cap (around 4000\,K).
Our estimated diameter of RS Cap is compatible with that reported in Richichi et
al. (\cite{Richichi1992}) from lunar
occultations and that reported in Dyck, van Belle \& Thomson (\cite{Dyck1998}) 
from IOTA observations, although the precision in our estimate is an order of
magnitude higher.

\subsubsection{A word of caution about the use of FINITO}

\begin{figure} 

\centering 

\includegraphics[width=9.0cm]{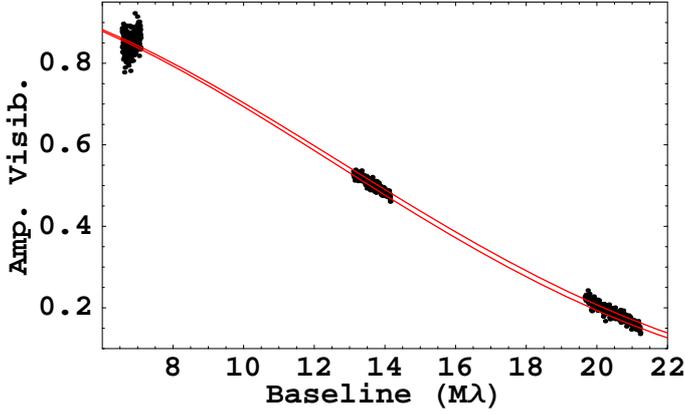}

\caption{Visibility amplitudes between 2.13 and 2.29\,$\mu$m as a function of baseline
length. Black, observations; red, model predictions with a uniform-disk diameter of 
$7.75 + 0.07$\,mas (lower line) and $7.75 - 0.07$\,mas (upper line).} 

\label{SizeComp} 

\end{figure}

We note that the use of FINITO in our observations may bias
the absolute amplitude calibration of the visibilities of each baseline, since
the percentage of lost fringes during the DIT is not taken into account in the
process of frame integration of each exposure. Any possible bias may be
different for calibrator and target, and translate into a different absolute
amplitude calibration for each star. As a result, our size estimate of RS Cap
might be (slightly) biased even after performing the calibration described in
Sect. \ref{LambdaCal}. However, the spectral resolution of our observations is 
much higher than that used in Richichi et al. (\cite{Richichi1992}) and Dyck, van 
Belle \& Thomson (\cite{Dyck1998}) (who, indeed, only measured the visibility at one
projected baseline).  The higher spectral resolution of AMBER allows us to 
precisely obtain the behavior of visibility amplitudes through relatively 
wide regions of the Fourier plane.  This sampling of visibility amplitudes through 
the observing band encodes information about the source size, which is independent of the
absolute calibration of the visibilities. In Fig. \ref{SizeComp}, we show the
visibility amplitudes in the continuum (i.e. blueward of 2.29\,$\mu$m) 
as a function of baseline length. In the figure, we also show 
the model predictions using a uniform-disk diameter of $7.75\pm0.07$\,mas. 
It can be seen that a disk model with this diameter satisfactorily fits the 
visibilities of all baselines through the whole spectral coverage in the continuum. 
This result gives us confidence about the amplitude calibration of the RS\,Cap 
visibilities.

\subsection{CO first overtone bands: $\lambda > 2.3$\,$\mu$m}
\label{CObands}

In Fig. \ref{RSCAPDISC}(c), there are two clear aspects in the region of 
CO first overtone bands ($\lambda > 2.3$\,$\mu$m) worth noticing. On 
the one hand, we see an increase of
$\sim10$\% in size for all the CO band heads, compared to the sizes in the
continuum. On the other hand, there is a linear trend of increasing size with
observing wavelength, beginning at 2.29\,$\mu$m. Owing to this latter effect, the
size in the continuum at 2.45\,$\mu$m is $\sim12$\% larger than the size in the
continuum for the wavelengths shorter than 2.29\,$\mu$m. 

Large size increases 
redward of 2.29\,$\mu$m have been observed in Mira stars 
(Woodruff et al. \cite{Woodruff2009}), which are interpreted as being caused by 
water-vapor extended zones. However, in those cases, the size variations reported 
are 3--4 times larger than that observed in RS Cap. 
In Sect. \ref{WaterSec}, we discuss how our model of water-vapor envelope can 
explain the observed increasing size in RS Cap redward of 2.29\,$\mu$m.

\section{Modeling}
\label{MODELING}

\subsection{{\tt MARCS} model atmosphere} 
\label{RESRSCAP}

We compared the observed size effects in RS Cap with different model
atmospheres computed with the {\tt MARCS} code (e.g. Gustafsson et al.
\cite{Gustafsson2008}).  The {\tt MARCS} code computes the hydrostatic
atmospheric structure in radiative and convective equilibrium for spherical
atmospheres with molecular and atomic lines taken into account using the opacity
sampling method.  Each model in spherical geometry (suitable for AGB stars) is
specified by the effective temperature ($T_{\mathrm{eff}}$), surface gravity
($\log{g}$), stellar mass ($M$), microturbulent velocity, and chemical
composition.

For $T_{\mathrm{eff}}$, we adopted 3200\,K, the nearest possible value 
in {\tt MARCS} models to our estimated 3160\,K.  
We also assume the ``moderately CN-cycled'' chemical composition (C/N = 
1.5 and $^{12}$C/$^{13}$C = 20), which is usually found in stars after 
the red giant branch, and [Fe/H] = 0.0.  
The stellar mass can be estimated by comparing with theoretical stellar 
evolutionary tracks, once $T_{\mathrm{eff}}$ and the luminosity are known.  
While the parallax of RS Cap is not precisely measured (the relative error 
is on the order of 70\%, so the error in the luminosity can be as large as 
an order of magnitude), 
Winters et al. (\cite{winters03}) estimated the luminosity of RS\,Cap to be 
$\log{L/L_{\odot}} = 3.82$, by applying the period-luminosity relation for 
semi-regular variables given by Feast (\cite{feast96})\footnote{Although the 
period-luminosity relation is a matter of debate, as Winters et al. 
(\cite{winters03}) mention.}. From Fig. 5 of Feast (\cite{feast96}), we estimated
an uncertainty of $\pm 0.25$ in the bolometric absolute magnitude of RS\,Cap, 
which translates into an error of up to 20--25\% in the luminosity, much smaller 
than that related to the Hipparcos parallax. The luminosity from Winters et al. 
(\cite{winters03}), together with the 
above $T_{\mathrm{eff}}$, places RS Cap close to the evolutionary track of a 
2\,M$_{\odot}$ star with Z = 0.02,
computed by F. Herwig\footnote{\tt http://astrowww.phys.uvic.ca/$\sim$fherwig}.  
Therefore, we adopted 2\,M$_{\odot}$ for RS Cap. 

We note that the luminosity estimated by Winters et al. (\cite{winters03})
for RS Cap is much higher than that corresponding to an RGB star with
2\,M$_{\odot}$ (which is only $\log{L/L_{\odot}} \sim 3$). Therefore, 
according to its luminosity, RS\,Cap would be clasified as an AGB star, 
as it is indeed assumed in the P-L relation by Feast (\cite{feast96}).
The luminosity estimated in this way for RS\,Cap is, thus, self-consistent 
in the framework of the P-L relation of semi-regular AGB stars.

From the luminosity and
$T_{\mathrm{eff}}$, we estimated a stellar radius of $\sim 260$\,R$_{\odot}$, 
which, combined with our angular-size estimate, translates into a distance
of about 310\,pc. Hence, from the adopted stellar mass of 
2\,M$_{\odot}$ and the estimated radius, we derive $\log{g} \sim -0.1$. We adopted 
$\log{g} = 0.0$ for the selection of the {\tt MARCS} models. 
For the microturbulent velocity, we used the {\tt MARCS} models with 2 and 
5\,km\,s$^{-1}$.  However, the effects of this parameter on the visibility 
are very small.


We downloaded the models with the above parameters from the {\tt MARCS}
website\footnote{\tt http://marcs.astro.uu.se}.  Using the temperature and
pressure stratifications of the model, we computed the monochromatic intensity
profiles and the synthetic spectrum for the spectral range of our AMBER observations
with a fine wavelength mesh of 1.4\,km\,s$^{-1}$, sampling several points over
each line profile.  The CO ($^{12}$C$^{16}$O, $^{12}$C$^{17}$O, $^{12}$C$^{18}$O,
and $^{13}$C$^{16}$O) and H$_2$O lines were taken from the line lists of
Goorvitch (\cite{goorvitch94}) and Partridge \& Schwenke (\cite{partridge97}),
respectively.  The monochromatic visibilities were computed from the Hankel
transform (i.e., Fourier transform of a circularly-symmetric distribution) of
the monochromatic intensity profiles (computed as described in 
Ohnaka et al. \cite{ohnaka06}) and then convolved to the spectral
resolution of our AMBER measurements.

In Quirrenbach et al. (\cite{Quirrenbach2001}), limb-darkened photospheric
models were used to generate equivalent disk sizes. These sizes were then
compared to the disk sizes fitted to the MkIII visibilities. In our case, this
approach would have had the disadvantage of being sensitive to scaling factors
between different baselines (i.e., to the absolute amplitude calibration) and
would have depended on the use of an intermediate model (a uniform disk) to compare
observations to models. To avoid those drawbacks, we chose a new approach
to the modeling of the RS Cap visibilities, which minimizes the effect of
possible calibration artifacts between baselines and, in addition, avoids the
use of an intermediate fit to a disk.
The synthetic visibilities obtained from the {\tt MARCS} models were computed at
the points of the Fourier plane corresponding to each exposure in our
observations. We then averaged the synthetic visibilities over all the
exposures, in the same way as we did with the real data. In this way, we directly 
compare the observed visibilities to the model predictions.

We show in Fig. \ref{MODWITHWATER} a comparison between observed visibilities
(black lines) and those computed from the {\tt MARCS} model using
$T_{\mathrm{eff}} = 3200$\,K, $M = 2$\,M$_{\odot}$, and $\log{g}=0$ (blue
lines). The match between observations and model predictions in
the continuum (i.e., for wavelengths shorter than $\sim2.3$\,$\mu$m) is 
very good.

The reduced $\chi^2$ for this model is $\chi^2_{\mathrm{red}} = 18.2$.
The relatively large depth of the features of the CO band heads cannot be
reproduced by the model. The discrepancy remains even 
if we lower the mass to values as low as $M=1$\,M$_{\odot}$ (models with higher 
masses predict more compact atmospheres and, therefore, smaller size effects 
in the CO band heads). Additionally, the observed visibility amplitudes at
wavelengths longer than $\sim$2.29\,$\mu$m are systematically lower than the
model predictions; this difference between model and observations is larger at
longer wavelengths.  In other words, the effective size of the star at
wavelengths longer than $\sim 2.29\,\mu$m is systematically larger than the
effective size derived from the {\tt MARCS} model.

\subsection{An additional water-vapor envelope}
\label{WaterSec}

An extra contribution must be added to the {\tt MARCS} models to fit the
visibilities in the observed spectral range.  We can reproduce the lower
visibility amplitudes (i.e., the larger effective sizes) beyond 2.29\,$\mu$m by
adding a contribution to the stellar opacity due to water vapor around the star.
The opacity corresponding to the wide absorption of water vapor centered 
at $\sim$2.7\,$\mu$m may account for the larger angular sizes observed at the longer
wavelengths.

We added a simple water-vapor model envelope to the {\tt MARCS}
pressure/density profiles, consisting of a narrow spherical shell with a radius
of 2 and a width of 0.1 (both in units of the stellar radius), a column density
of $10^{21}$\,cm$^{-2}$, and a temperature of 1500\,K.  Using these parameters,
we are able to fit the amplitude visibilities at $\lambda > 2.29\,\mu$m, as we
show in Fig \ref{MODWITHWATER} (red lines). With this model, we obtain
$\chi^2_{\mathrm{red}} = 8.1$. This large reduced $\chi^2$ is partially
due to the discrepancy in the CO bands, which is discussed in Sect. \ref{COExtSec}. 
Our model follows the general trend of decreasing visibility 
amplitudes for wavelengths longer than 2.3\,$\mu$m, although underestimates 
the visibility amplitudes at wavelengths longer than 2.4\,$\mu$m. Changing 
the column density by a factor of ten
results in an increase of $\chi^2_{\mathrm{red}}$ by a factor $2-3$; a slight
modification of the water-vapor temperature ($\sim15$\%) increases
$\chi^2_{\mathrm{red}}$ by a factor $\sim2$; and changing the size of the shell
by a factor of two increases the $\chi^2_{\mathrm{red}}$ by a factor $\sim2$. 

Although the water-vapor envelope enhances the contrast of the CO
features in the visibilities, thus improving the fit also in the CO band heads, 
the observed depths in the band heads are still larger than those of the 
model for some band heads ($\sim 50$\% for the band head at 2.29\,$\mu$m). 
Furthermore, increasing
(decreasing) the stellar mass results in smaller (larger) depths of the
visibility amplitudes in the CO band heads, but decreasing the mass to a
value as low as 1\,M$_{\odot}$ does not translate into CO visibility
amplitudes comparable to those of the observations, even after the addition of
the water envelope (the $\chi^2_{\mathrm{red}}$ using 1\,M$_\odot$ is only 7.2\%
lower than that using 2\,M$_\odot$). 

\begin{figure} 

\centering 

\includegraphics[width=9.5cm]{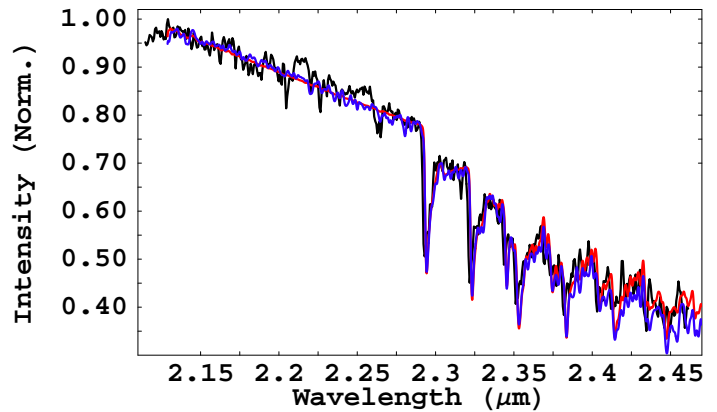}

\caption{Calibrated spectrum of RS Cap (black), synthetic {\tt MARCS} spectrum (blue),
and synthetic spectrum with the additional water-vapor envelope (red).} 

\label{SpectrumFit} 

\end{figure}

In Fig. \ref{SpectrumFit}, we show the calibrated spectrum of RS Cap (black) and, 
superimposed, the synthetic spectra obtained from the {\tt MARCS} model only 
(blue) and the {\tt MARCS} model with an additional water-vapor envelope (red). 
Both model spectra are very similar, 
and only differ for the longest wavelengths, where the model with water envelope
predicts a slightly lower flux. Since the fit of both models to the observed spectrum is of 
similar quality, it is difficult to extract any additional conclusion about the water-vapor 
envelope from the spectrum alone. 

\subsubsection{Water-vapor envelopes in AGBs. The problem of RS Cap}

The presence of dense water vapor envelopes in semi-regular or irregular 
(i.e., non-Mira-type) AGB stars was first revealed 
by Tsuji et al. (\cite{tsuji97}), who unmasked the 2.7\,$\mu$m water 
vapor band originating in the dense molecular layers extending to $\sim$2 
stellar radii.   
On the basis of infrared interferometric observations, Mennesson et al. 
(\cite{Mennesson2002}) found an increase in the angular size of
30--60\% between the $K$ and $L^{\prime}$ band in five semi-regular 
AGB stars, which the authors interpreted 
as being due to the emission from the extended molecular layers.  However, 
they used filters which are incapable of spectrally resolving 
the water vapor bands.  
Our AMBER measurements are the first study to {\it spatially and spectrally} 
resolve the water vapor emission from the dense molecular layers for a 
semi-regular AGB star.

We note that a similar increase in the angular diameter 
redward of 2.3\,$\mu$m is also found in Mira stars.  
Based on $K$-band interferometry using narrow-band filters 
with a spectral resolution of $\sim$20, 
Perrin et al. (\cite{Perrin2004}) showed that the angular diameters of 
Mira stars are larger at 2.0 and 2.4\,$\mu$m than at 2.2\,$\mu$m.  
Wittkowski et al. (\cite{wittkowski08}) observed 
the Mira star S Ori from 1.29 to 2.32\,$\mu$m using AMBER with a spectral 
resolution of 35 and found that the angular size increases redward 
of 2.2\,$\mu$m.  
The wavelength dependence of the angular size of Mira 
stars from the near- to mid-infrared is primarily governed by the opacity 
of water vapor, which is abundant in the atmosphere of Mira stars 
(e.g., Ohnaka \cite{Ohnaka2004}; Wittkowski et al. \cite{wittkowski08}).  
The model by Tej et al. (\cite{tej03}) shows that a shell with an enhanced 
water vapor abundance forms behind a shock front propagating outward in the 
extended atmosphere of Mira stars.  
Comparison between infrared interferometric observations of Mira stars and 
dynamical models lends support to this picture 
(e.g., Ohnaka et al. \cite{ohnaka06}; Wittkowski et al. \cite{wittkowski07}).

The shock wave in Miras is generated by the periodic, large-amplitude 
pulsation.  Therefore, the shock in non-Mira stars with much 
smaller variability amplitude is expected to be much weaker.  
However, despite the striking difference in the variability amplitude 
($\Delta V$ = 6--9 and 0.5--2 for Miras and RS Cap, respectively), 
the semi-regular AGB star RS Cap produces a warm water vapor envelope 
similar to that of Miras.  
The column density, temperature, and radius of the water vapor envelope derived 
for RS Cap are similar to those derived for Mira stars 
(Ohnaka \cite{Ohnaka2004}). Not only the properties of the water-vapor 
envelope, but also the mass-loss rate of RS Cap 
($\sim \! 1 \times 10^{-6}$\,M$_{\odot}$ yr$^{-1}$) is comparable to 
that of Miras (e.g., $5 \times 10^{-7}$\,M$_{\odot}$ yr$^{-1}$ for 
S Ori) (estimates from Winters et al. \cite{winters03}).  

The dynamical models with small-amplitude pulsations, reported in 
Winters et al. (\cite{winters03}), produce outflows slower than 5\,km s$^ {-1}$ with 
mass-loss rates lower than $3 \times 10^{-7}$\,M$_{\odot}$ yr$^{-1}$. However, 
the expansion velocity measured for RS Cap is $\sim$10\,km s$^{-1}$ 
(Winters et al. \cite{winters03}).  
Therefore, it is not clear whether this kind of small-amplitude pulsation 
model applies to RS Cap.  
At the moment, there is no satisfactory explanation for the origin of 
the warm water envelope in RS Cap.

\subsection{An additional CO envelope}
\label{COExtSec}

\begin{figure} 

\centering 

\includegraphics[width=9.5cm]{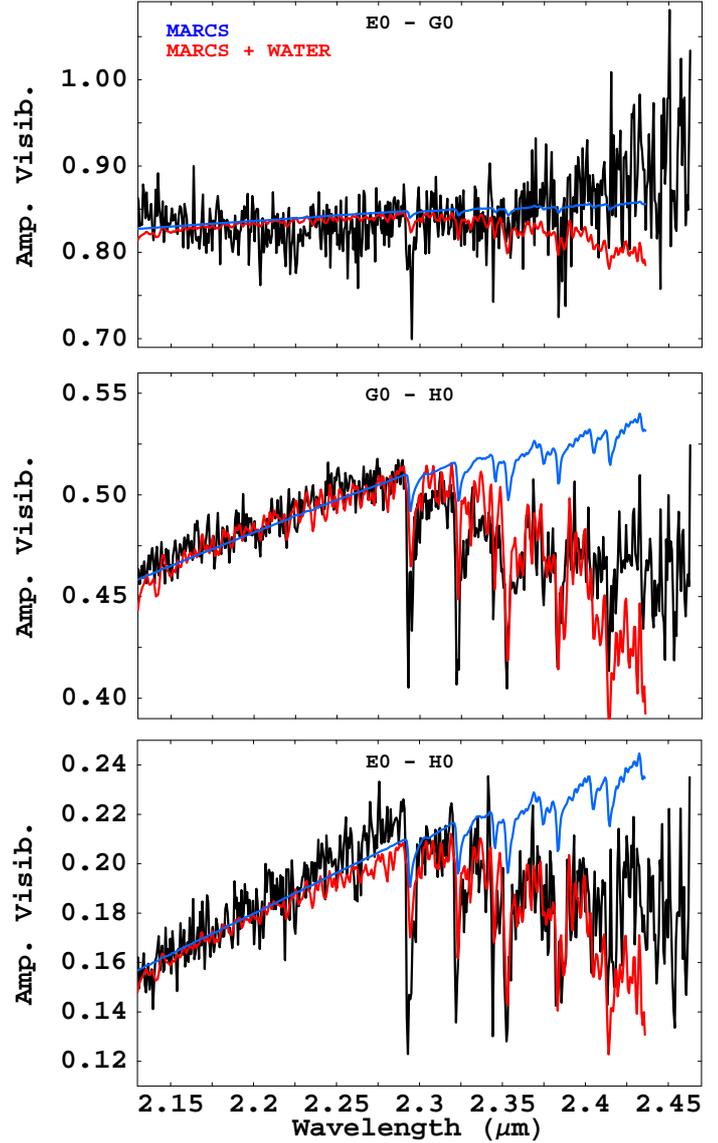}

\caption{Black: AMBER visibilities of RS Cap. Blue: model visibilities computed
using a {\tt MARCS} pressure-density profile (see text).  Red: model
visibilities computed using the {\tt MARCS} pressure-density profile and an
additional water-vapor envelope (see text).} 

\label{MODWITHWATER} 

\end{figure}

It can be seen in Fig. \ref{MODWITHWATER} that the observed visibility amplitudes at 
the CO band heads are much lower than the model predictions, either with or 
without a water-vapor envelope. This discrepancy is especially large for the 
shortest baseline (i.e. $E0-G0$) at the 2.29$\mu$m band head. The low visibility 
amplitudes at the CO band 
heads provide evidence of an extended envelope of CO around the star. For instance, if 
$\sim$20\% of the emission in the CO band head at 2.29\,$\mu$m were to originate from 
a very extended envelope of, say, $\sim$20\,mas radius or larger, it would map 
into a decrease in the visibility amplitude that would roughly match the 
observations in that part of the spectrum. Unfortunately, we do not have enough 
data to be able to characterize well this possible extra component of CO 
absorption. 


\section{Conclusions} \label{CONCLUSIONS}

We have observed the AGB star RS Cap with VLTI/AMBER in the K band with
medium-resolution mode (506 channels between 2.13 and 2.47\,$\mu$m). We 
have estimated
a Rosseland diameter of $7.95\pm0.07$\,mas in the continuum, which translates
into an effective temperature of 3160$\pm$160\,K.  The apparent size of the star
increases monotonically by $\sim12$\% between 2.29 and 2.47\,$\mu$m.  We have
detected lower than expected visibility amplitudes in all the CO band heads
observed. These lower amplitudes translate into larger apparent sizes of the
star in the CO band heads. 

Using the pressure/density model profiles obtained with the {\tt MARCS} code, we
have been able to generate synthetic visibilities and compare them directly to our
AMBER observations. The fit is rather good, although we are unable to
reproduce the low visibility amplitudes in the CO band heads using a mass for RS
Cap estimated from the theoretical models of stellar evolution. In this sense,
our situation resembles that of Quirrenbach et al. (\cite{Quirrenbach2001}), who
observed interferometrically a sample of cool-giant stars in the TiO absorption
band at 712\,nm.  The discrepancy between models and observations in the CO band
heads might be resolved either by using a much lower mass for RS Cap (below
1\,M$_{\odot}$, thus in contradiction with the stellar-evolution models) or by
additional unmodeled effects in the base of the stellar wind (a transition zone
with prominent absorption in the CO bands).

To fit the observations at wavelengths longer than 2.29\,$\mu$m,
we have found that an {\em ad hoc} narrow spherical water-vapor envelope around the
star, similar to the models used in Perrin et al. (\cite{Perrin2004}), must be 
added. We have modeled 
this envelope with a temperature of 1500\,K, a size twice
that of the star, a width of 0.1 times the stellar radius, and a column density
of $10^{21}$\,cm$^{-2}$.

Finally, there is a hint of an extended CO envelope around the star,
based on the low visibility amplitudes at the CO band heads in the shortest baseline.

\acknowledgements{ IMV is a fellow of the Alexander von Humboldt Foundation in
Germany.  This research has made use of the SIMBAD database, operated at CDS,
Strasbourg, France. Partial support from Spanish grants AYA-2006-14986-C02, 
AYA2009-13036-C02-02, and Prometeo 2009/104 is acknowledged.  }

\end{document}